\documentclass[prl,showpacs,superscriptaddress,amsmath,amssymb,twocolumn,aps]{revtex4}
\usepackage{graphicx}
\usepackage{dcolumn}
\usepackage{bm}

\begin{document}

\title{Experimental observation of Dirac-like surface states and topological phase transition in Pb$_{1-x}$Sn$_x$Te(111) films}

\author{Chenhui Yan}
\affiliation{State Key Laboratory of Low-Dimensional Quantum Physics, Department of Physics, Tsinghua University, Beijing 100084, China}
\affiliation{Collaborative Innovation Center of Quantum Matter, Beijing 100084, China}
\affiliation{Shenyang National Laboratory for Materials Science, Institute of Metal Research, Chinese Academy of Sciences, Shenyang 110016, China}
\author{Junwei Liu}
\affiliation{State Key Laboratory of Low-Dimensional Quantum Physics, Department of Physics, Tsinghua University, Beijing 100084, China}
\affiliation{Collaborative Innovation Center of Quantum Matter, Beijing 100084, China}
\author{Yunyi Zang}
\affiliation{State Key Laboratory of Low-Dimensional Quantum Physics, Department of Physics, Tsinghua University, Beijing 100084, China}
\affiliation{Collaborative Innovation Center of Quantum Matter, Beijing 100084, China}
\author{Jianfeng Wang}
\affiliation{State Key Laboratory of Low-Dimensional Quantum Physics, Department of Physics, Tsinghua University, Beijing 100084, China}
\affiliation{Collaborative Innovation Center of Quantum Matter, Beijing 100084, China}
\author{Zhenyu Wang}
\affiliation{State Key Laboratory of Low-Dimensional Quantum Physics, Department of Physics, Tsinghua University, Beijing 100084, China}
\affiliation{Collaborative Innovation Center of Quantum Matter, Beijing 100084, China}
\author{Peng Wang}
\affiliation{State Key Laboratory of Low-Dimensional Quantum Physics, Department of Physics, Tsinghua University, Beijing 100084, China}
\affiliation{Collaborative Innovation Center of Quantum Matter, Beijing 100084, China}
\author{Zhi-Dong Zhang}
\affiliation{Shenyang National Laboratory for Materials Science, Institute of Metal Research, Chinese Academy of Sciences, Shenyang 110016, China}
\author{Lili Wang}
\affiliation{State Key Laboratory of Low-Dimensional Quantum Physics, Department of Physics, Tsinghua University, Beijing 100084, China}
\affiliation{Collaborative Innovation Center of Quantum Matter, Beijing 100084, China}
\author{Xucun Ma}
\affiliation{State Key Laboratory of Low-Dimensional Quantum Physics, Department of Physics, Tsinghua University, Beijing 100084, China}
\affiliation{Collaborative Innovation Center of Quantum Matter, Beijing 100084, China}
\author{Shuaihua Ji}
\affiliation{State Key Laboratory of Low-Dimensional Quantum Physics, Department of Physics, Tsinghua University, Beijing 100084, China}
\affiliation{Collaborative Innovation Center of Quantum Matter, Beijing 100084, China}
\author{Ke He}
\affiliation{State Key Laboratory of Low-Dimensional Quantum Physics, Department of Physics, Tsinghua University, Beijing 100084, China}
\affiliation{Collaborative Innovation Center of Quantum Matter, Beijing 100084, China}
\author{Liang Fu}
\affiliation{Department of Physics, Massachusetts Institute of Technology, Cambridge, Massachusetts 02139, USA}
\author{Wenhui Duan}
\affiliation{State Key Laboratory of Low-Dimensional Quantum Physics, Department of Physics, Tsinghua University, Beijing 100084, China}
\affiliation{Collaborative Innovation Center of Quantum Matter, Beijing 100084, China}
\author{Qi-Kun Xue}
\email{qkxue@mail.tsinghua.edu.cn} 
\affiliation{State Key Laboratory of Low-Dimensional Quantum Physics, Department of Physics, Tsinghua University, Beijing 100084, China}
\affiliation{Collaborative Innovation Center of Quantum Matter, Beijing 100084, China}
\author{Xi Chen}
\email{xc@mail.tsinghua.edu.cn}
\affiliation{State Key Laboratory of Low-Dimensional Quantum Physics, Department of Physics, Tsinghua University, Beijing 100084, China}
\affiliation{Collaborative Innovation Center of Quantum Matter, Beijing 100084, China}

\date{\today}

\begin{abstract}
The surface of a topological crystalline insulator (TCI) carries an even number of Dirac cones protected by crystalline symmetry. 
We epitaxially grew high quality Pb$_{1-x}$Sn$_x$Te(111) films and investigated the TCI phase 
by in-situ angle-resolved photoemission spectroscopy. 
Pb$_{1-x}$Sn$_x$Te(111) films undergo a topological phase transition from trivial insulator to TCI via increasing the Sn/Pb ratio, 
accompanied by a crossover from n-type to p-type doping. 
In addition, a hybridization gap is opened in the surface states when the thickness of film is reduced to the two-dimensional limit.
The work demonstrates an approach to manipulating the topological properties of TCI, 
which is of importance for future fundamental research and applications based on TCI.
\end{abstract}


\maketitle

Recently topological classification of quantum matter has been extended to a new class of matter, namely, 
topological crystalline insulators (TCI) \cite{fu11}. A TCI consists of a bulk gap and an even number of robust  surface Dirac cones. 
Different from the well-known topological insulators (TIs) \cite{zhang10,kane10,zhang11}, 
the topological surface states of TCI are protected by crystalline symmetry instead of time-reversal symmetry (TRS). 
Compared with TI, TCI offers a platform exploring broader topology-related phenomena, for example,  
spin-filtered edge states with electrically tunable gap \cite{fu13a} 
and large Chern number quantum anomalous Hall phases \cite{bernevig13}.
First-principles calculation demonstrates that TCI can be realized in SnTe \cite{fu12}. 
In contrast, PbTe, another IV-VI semiconductor with similar structure as SnTe, is topologically trivial. 
Therefore, the Pb$_{1-x}$Sn$_x$Te compound is expected to undergo a topological phase transition at certain value of $x$. 
Here we report the observation of topological surface states and the phase transition from a trivial insulator 
to a TCI in high quality Pb$_{1-x}$Sn$_x$Te(111) films prepared by molecular beam epitaxy (MBE). 
In addition, we observed that the (111) surface of topological Pb$_{1-x}$Sn$_x$Te harbors Dirac cones at the so-called time-reversal-invariant momenta
(TRIM) \cite{fu13b}.  Its observation has been a challenge owing to the difficulty in sample preparation. 
The thin film also enables us to investigate the thickness-dependent band structure
of TCI. A hybridization gap is clearly seen when the thickness is reduced to a few nanometers. 

Figures 1(a) and 1(b) show the rock-salt structure of Sn(Pb)Te and the Brillouin zone of this narrow-gap semiconductor.
The fundamental band gaps are located at four equivalent L points in the Brillouin zone. 
The spin-orbit coupling  can be altered by the substitution of Sn for Pb in Pb$_{1-x}$Sn$_x$Te.
As a result, the band gaps close at a critical composition and re-open with increasing Pb content \cite{strauss66,buczko13}.
The order of conduction and valence bands at the L points is inverted at the critical composition.  
The band inversion changes the mirror Chern number $n_M$ \cite{kane07} from -2 (topological nontrivial) to 0 (topological trivial), 
giving rise to the topological phase transition as schematically shown in Fig. 1(c).

\begin{figure*}[t]
        \includegraphics[width=5.5in]{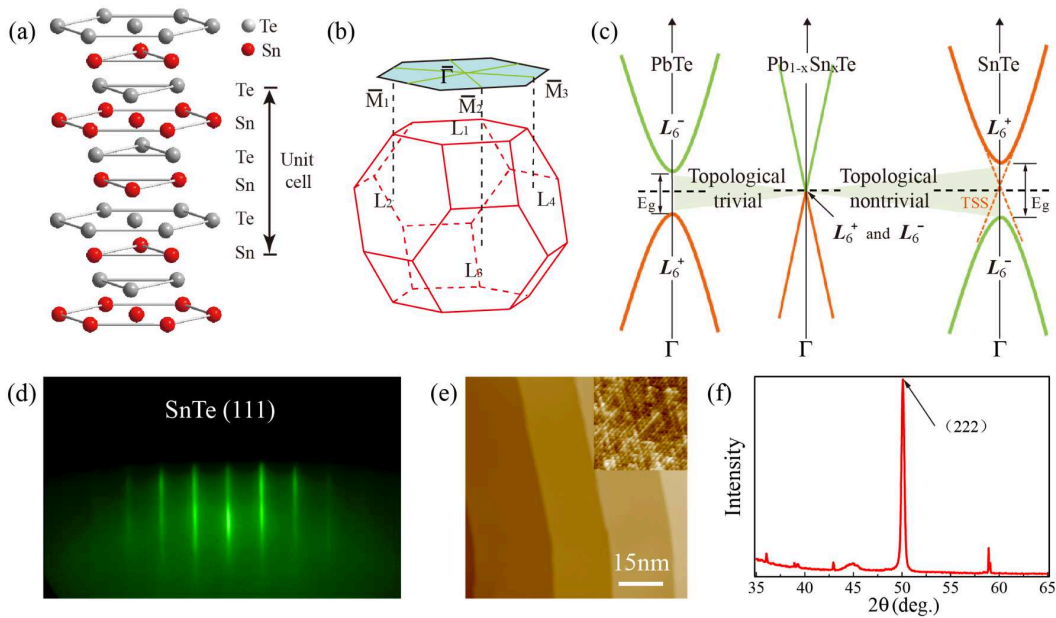}
        \caption{\label{fig1} MBE growth of Pb$_{1-x}$Sn$_x$Te(111) films. 
The growth dynamics is very similar to that of Bi$_2$Te$_3$ and Bi$_2$Se$_3$ \cite{xue10a,xue11,xue10b}. 
High-quality single crystalline  thin film is achieved under Te-rich condition (flux ratio Te/Sn$>$5) 
and T$_\text{Sn(Pb)}$$>$T$_\text{sub}$$>$T$_\text{Te}$, where T$_\text{Sn(Pb)}$, T$_\text{sub}$ and T$_\text{Te}$ 
are the temperatures of Sn(Pb)-cell, substrate and Te-cell, respectively. The growth rate is typically $\sim$0.25 monolayer/sec (ML/s) 
at T$_\text{Sn}$=1020 $^\circ$C, T$_\text{sub}$=310 $^\circ$C and T$_\text{Te}$=300 $^\circ$C. 
Bi$_2$Te$_3$ buffer layer is 7 nm thick. (a) Crystal structure of Sn(Pb)Te along the [111] direction. 
(b) The bulk and the projected (111) Brillouin zones of Pb$_{1-x}$Sn$_x$Te. 
(c) Schematic illustration of topological phase transition in Pb$_{1-x}$Sn$_x$Te system. TSS stands for topological surface states.
(d) RHEED pattern along the [1$\bar{1}$0] ($\bar{\Gamma}-\bar{\text{K}}$) direction of SnTe(111) film of 30 nm thick. 
(e) STM image of SnTe(111) film acquired at 77 K. The inset is the atomically resolved image. (f) XRD pattern of SnTe film. 
Only (2,2,2) Bragg peak is clearly seen. }
\end{figure*}

The nontrivial surface states of topological Pb$_{1-x}$Sn$_x$Te  exist on the high-symmetry surfaces, 
such as \{001\}, \{110\} and \{111\} that preserve the mirror symmetry with respect to the \{1$\bar{1}$0\} planes. 
Here the notation \{hkl\} refers to the (hkl) plane and all those equivalent ones under symmetry transformation.
Depending on the surface orientation, there are two types of surface states with qualitatively different properties: 
they are either located at TRIM or non-TRIM. More specifically, the (111) surface states of topological Pb$_{1-x}$Sn$_x$Te 
consist of totally four Dirac cones centered at 
TRIM $\bar{\Gamma}$ and $\bar{\text{M}}$ points, respectively (see Fig. 1(b) for the (111) surface Brillouin zones). 
So far most of the experiments \cite{ando12,story12,hasan12,chen14,ando13a,madhavan13,ando13b,tjernberg13} 
on Pb$_{1-x}$Sn$_x$Te have been performed on the (001) surface, which is the natural cleavage plane of IV-VI semiconductors. 
The (111) surface is a polar surface and difficult to obtain in single crystal growth. 
To meet the challenge, we prepared  Pb$_{1-x}$Sn$_x$Te(111) thin films by using MBE. 
The lattice constant of SnTe (PbTe) along the [111] crystallographic direction is 1.82 (1.86) {\AA} 
and that on the hexagonal (111) plane is 4.45 (4.56) {\AA}. The in-plane lattice constant of Pb$_{1-x}$Sn$_x$Te(111) 
is very close to that of Bi$_2$Te$_3$. We therefore choose Bi$_2$Te$_3$ thin film grown on Si(111) as the substrate 
in epitaxial growth \cite{ando13c}.

The experiments were performed in an ultrahigh vacuum system that consists of an MBE growth chamber, 
a low temperature scanning tunneling microscope (STM) (omicron) and an angle-resolved photoemission spectrometer (VG-scienta) 
with a base pressure better than $1\times10^{-10}$ mbar. The Si(111)-7$\times$7 substrate was prepared by multi-cycle flashing 
to 1200 $^\circ$C. High purity Bi (99.999\%), Te (99.9999\%), Sn (99.9999\%) and Pb (99.999\%) were evaporated from standard Knudsen cells. 
Real-time reflection high-energy electron diffraction (RHEED) was used to monitor the film growth and 
calibrate the growth rate according to the intensity oscillation of (0,0) diffraction. 
The film morphology was characterized by STM at 77 K with platinum-iridium tips. 
In the angle-resolved photoemission spectroscopy (ARPES) measurement, 
samples were kept at 77 K and a Scienta R4000 analyzer was used to collect the photoelectrons 
excited by He-I light source of 21.2 eV. The energy and angular resolution were better than 20 meV and 0.2$^\circ$, respectively. 

The atomically flat surface morphology of the as-grown films is revealed by the sharp streak of 1$\times$1 RHEED pattern 
along the [1$\bar{1}$0] direction in Fig. 1(d). When the incidence direction of electron beam in RHEED was turned to the [11$\bar{2}$] direction, 
another sharp 1$\times$1 RHEED pattern was observed (not shown in the figure), 
suggesting that the film is (111)-oriented and there is no surface reconstruction. 
In the STM image [Fig. 1(e)], the step height on the film is $\sim$3.6 {\AA}, corresponding to the thickness of Sn(Pb)-Te 
double layer along the (111) direction. The atomic-resolution STM image in the inset of Fig. 1(e) illustrates the 
hexagonal in-plane lattice structure of the Pb$_{1-x}$Sn$_x$Te(111) surface. The high crystal quality and (111) surface orientation
are also confirmed by the x-ray diffraction (XRD) pattern of the film [Fig. 1(f)]. 
Furthermore, density functional theory calculations (see supplemental material \cite{supplemental}) 
and previous work \cite{zhang13} suggest that the SnTe (111) surface is terminated by Te atoms. 
If the (111) surface were Sn-terminated, it should reconstruct to diminish the surface energy. 

\begin{figure}[h]
        \includegraphics[width=3.25in]{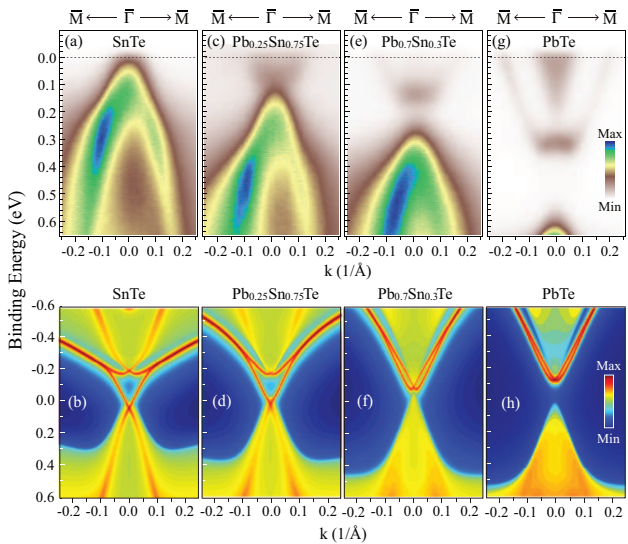}
        \caption{\label{fig2} 
Electronic structure of Pb$_{1-x}$Sn$_x$Te(111) film (30 nm thick) in the vicinity of $\bar{\Gamma}$ point. 
(a) and (b)   ARPES and TB calculation of SnTe.     (c) and (d) ARPES and TB calculation of Pb$_{0.25}$Sn$_{0.75}$Te.
(e) and (f)  ARPES and TB calculation of Pb$_{0.7}$Sn$_{0.3}$Te. (g) and (h) ARPES and TB calculation of PbTe.     
        }
\end{figure}

Figure 2(a) exhibits the ARPES spectra of SnTe film in the vicinity of $\bar{\Gamma}$ point. 
The observed electronic structure is identified as the valence band based on the tight-binding (TB) calculation \cite{ho86} [Fig. 2(b)] 
using Green's function method \cite{rubio85,dai08} (see supplemental material \cite{supplemental}).
The band bending effect \cite{hofmann10,hofmann12}, which commonly exists on the surface of narrow band semiconductors,
has been considered in the calculation. 
The lower Dirac cone of the topological surface states merges into the valence band and can not be distinguished.

The Fermi level in Fig. 2(a) intersects the bulk valence band due to the p-type Sn vacancy as the dominant dopant in SnTe 
\cite{kita09,siebenmann65,lyubchenko05}. 
It is therefore difficult to access the Dirac point of the topological surface states of SnTe in ARPES measurement 
\cite{ando12,hasan12,chen14,ando13a}. 
The Fermi level can be tuned by changing the Pb/Sn ratio in Pb$_{1-x}$Sn$_x$Te \cite{hasan12,chen14,ando13a}. 
With increasing Pb content, more and more n-type Te vacancies appear and gradually compensate the hole carriers. 
At low Pb/Sn ratio [Fig. 2(c)],  a Dirac-like dispersion becomes visible at $\bar{\Gamma}$ point. The Dirac point is located at about 130 meV
below the Fermi level and very close to the top of valence band.  TB calculation [Fig. 2(d)] ascribes this linear dispersion to the 
topological surface states.

ARPES measurement and TB calculation [Figs. 2(e) to 2(h)] indicate that 
Pb$_{1-x}$Sn$_x$Te at high Pb content is topologically trivial. 
The topological phase transition occurs when $x$ is between $0.3$ and $0.4$.
In the case of pure PbTe [Fig. 2(g)], the film is heavily n-doped and a bulk electron pocket is clearly visible at the Fermi level. 
Inside the bulk energy gap of 280 meV, no state with Dirac-like dispersion is observed. Instead, a W-shape band shows up.
Based on the TB calculation with band bending [Fig. 2(h)], we attribute this W-shape band to the trivial surface state with Rashba splitting.
The (111) plane of IV-VI semiconductors is a polar surface with dangling bonds, leading to the observed surface states. 
The band bending effect \cite{hofmann10,hofmann12} and spin-obit coupling give rise to 
the  strong Rashba splitting of the surface states at the $\bar{\Gamma}$ point.
The two sub-bands with different spins shift in opposite direction along the k axis and degenerate at the $\bar{\Gamma}$ point, 
resulting in the W-shape dispersion. The Rashba-type band persists in the non-topological Pb$_{1-x}$Sn$_x$Te compound
(for example, see Figs. 2(e) and 2(f) for Pb$_{0.7}$Sn$_{0.3}$Te), and apparently evolves into the topological surface states after
the topological phase transition. 

\begin{figure}[h]
        \includegraphics[width=3.25in]{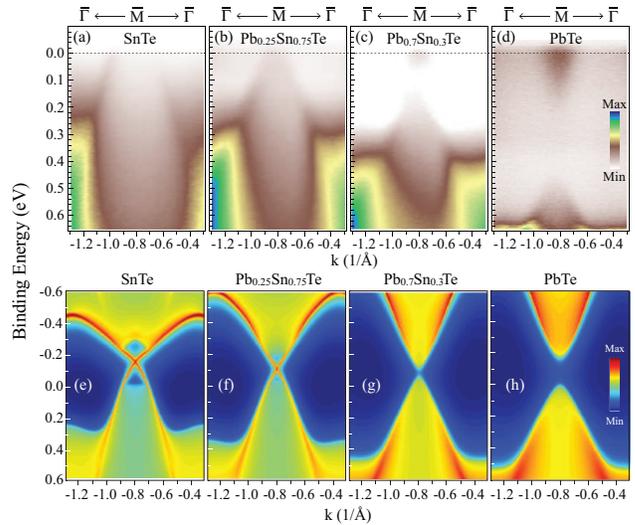}
        \caption{\label{fig3} 
        Electronic structure of Pb$_{1-x}$Sn$_x$Te(111) film (30 nm thick) in the vicinity of $\bar{\text{M}}$ point. 
(a) to (d) ARPES of SnTe, Pb$_{0.25}$Sn$_{0.75}$Te, Pb$_{0.7}$Sn$_{0.3}$Te and PbTe. (e)-(h) The corresponding
TB calculation, which  agrees well with the observed spectra.
        }
\end{figure}

One single Dirac cone at the $\bar{\Gamma}$ point is not enough to establish the notion of TCI. 
The topological Pb$_{1-x}$Sn$_x$Te is distinct from the $Z_2$ topological insulator and should carry an even number of Dirac points 
on the high-symmetry crystal surfaces.  Further evidence for TCI emerges at the $\bar{\text{M}}$ points. 
Figures 3(a) to 3(d) exhibit the ARPES intensity maps of Pb$_{1-x}$Sn$_x$Te with various compositions at $\bar{\text{M}}$ points. 
At low content of Pb, a Dirac-like band with linear dispersion is clearly resolved. 
The energy of the Dirac point is estimated to be 180 meV (SnTe) and 36 meV (Pb$_{0.25}$Sn$_{0.75}$Te) 
above the Fermi level by linear extrapolation. 
Together with the topological surface states at $\bar{\Gamma}$ point, 
there are totally four Dirac cones on the (111) surface as predicted by theory \cite{fu13b}.
With increasing content of Pb, Pb$_{1-x}$Sn$_x$Te becomes topologically trivial and 
there is no more topological surface state inside the bulk energy gap [Figs. 3(c) and 3(d)]. 
In addition, the bulk gap of PbTe at $\bar{\text{M}}$ point is larger than that of Pb$_{0.7}$Sn$_{0.3}$Te, which is consistent with 
the scenario that the gap closes at the critical composition and gradually re-opens with increasing Pb content 
after the topological phase transition. 

Different from previous band calculations \cite{fu13b,buczko13}, 
the location of Dirac point at $\bar{\text{M}}$ [Fig. 3(b)] is much higher than that at $\bar{\Gamma}$ [Fig. 2(c)].
The discrepancy can be resolved by taking into account the effect of band bending as shown 
in the TB calculation [Figs. 3(e) to 3(h)]. 

 \begin{figure}[h]
        \includegraphics[width=3in]{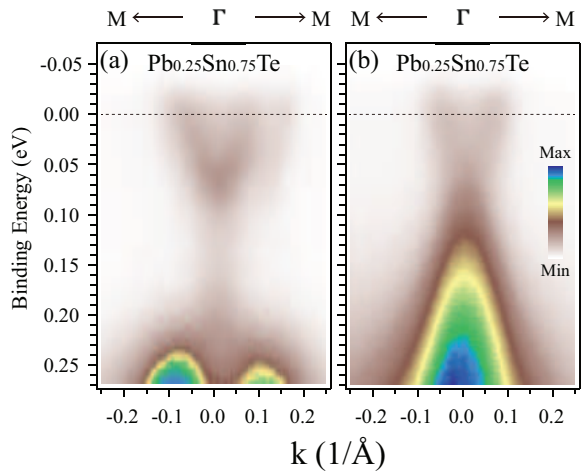}
        \caption{\label{fig4}  
ARPES of (a) 1.1 nm and (b) 30 nm   Pb$_{0.25}$Sn$_{0.75}$Te(111) films.}
\end{figure}

Finally, if the film thickness becomes thin enough, the coupling of topological states from the opposite surfaces 
should open an energy gap in the spectra
\cite{xue10b}. The gap opening ($\sim$ 170 meV) for a 1.1 nm Pb$_{0.25}$Sn$_{0.75}$Te film is clearly demonstrated in Fig. 4(a).
The vertically nondispersive feature between the upper and lower cones 
may stem from the enhanced many-body electronic interaction \cite{qian13}
or film inhomogeneity. 
For comparison, Fig. 4(b) shows the full Dirac cone without energy gap for a thick film ($\sim$ 30 nm). 

In summary, we have investigated the electronic structure of Pb$_{1-x}$Sn$_x$Te(111) thin film by ARPES 
and TB calculation. The MBE film exhibits a topological phase transition with increasing Pb content 
and an even number of Dirac cones in the TCI phase. 
Together with the property of high electronic mobility and large mean-free path for the (111)-oriented IV-VI semiconductors
\cite{bauer93,bauer06,bauer10},
the quasi-two dimensional thin film with exotic surface states 
paves the road for searching new quantum phases in TCI near the topological phase transition
and achieving high-efficiency electrical spin manipulation.

The work was financially supported by NSFC (11025419 and 51331006) and MOST (2011CB921904).

\end{document}